\documentclass[showpacs,aps,prl,twocolumn,floatfix]{revtex4}

\usepackage{amssymb,amsfonts}
\usepackage{times}
\usepackage{bm}
\usepackage{amsmath}
\usepackage{graphicx}

\newcommand{\s}{\sum\limits}

\newcommand{\pa}{\partial}
\newcommand{\il}{\int\limits}
\newcommand{\be}{\begin{equation}}
\newcommand{\e}{\end{equation}}
\newcommand{\beml}{\begin{subequations}}
\newcommand{\eml}{\end{subequations}}
\newcommand{\beq}{\begin{eqnarray}}
\newcommand{\eq}{\end{eqnarray}}
\newcommand{\ba}{\begin{array}}
\newcommand{\ea}{\end{array}}
\newcommand{\lt}{\left}
\newcommand{\rt}{\right}
\newcommand{\n}{\nonumber}
\newcommand{\la}{\langle}
\newcommand{\ra}{\rangle}
\newcommand{\ep}{\varepsilon}

\newcommand{\h}{^\dagger}
\newcommand{\up}{\uparrow}
\newcommand{\down}{\downarrow}

\DeclareMathOperator{\arcth}{arctanh}

\begin{document}
\title{Interaction-induced Renormalization of Andreev Reflection}
\author{M. Titov}
\author{M. M\"uller}
\author{W. Belzig}
\affiliation{Department of Physics, University of Konstanz, D--78457 Konstanz, Germany}
\date{July 31, 2006}
\begin{abstract}
  We analyze the charge transport between a one-dimensional weakly
  interacting electron gas and a superconductor within the scaling
  approach in the basis of scattering states. We derive the renormalization 
  group equations, which fully account for the intrinsic energy dependence 
  due to Andreev reflection. A strong renormalization of the 
  corresponding reflection phase is predicted even for a perfectly transparent metal-superconductor
  interface. The interaction-induced suppression of the Andreev conductance  
  is shown to be highly sensitive to the normal state resistance, 
  providing a possible explanation of
  experiments with carbon-nanotube/superconductor junctions by
  Morpurgo {\it et al.}  [Science {\bf 286}, 263 (2001)].
\end{abstract}
\maketitle

The superconducting proximity effect is a well-known phenomenon, which
has motivated a number of theoretical and experimental studies since
the middle of the last century.  The low-energy physics of the
proximity effect is described by Andreev reflection processes \cite{A}
at the boundary between a normal metal (N) and a superconductor (S).
In this process an electron-like quasiparticle in N
is reflected from the NS boundary as a hole, thus
transferring a double electron charge $2e$ into S. 
The probability of such an event tends to unity in the
case of an ideal NS interface provided the quasiparticle energy $\ep$ is
below the superconducting gap $\Delta$. Normal reflection takes place
at non-ideal interfaces due to
the Fermi-energy mismatch in the superconducting and normal-metal
materials or due to interface impurities.

In the elastic theory of electron transport, the NS boundary
is characterized by energy-dependent quantum-mechanical amplitudes: $r^A(\ep)$ 
for Andreev reflection and $r^N(\ep)$ for normal reflection.
The differential Andreev conductance of an NS junction measured at the voltage bias $V$
is given by \cite{BTK}
\be
\label{BTK}
\frac{\pa I}{\pa V}=\frac{2 e^2}{h} \int \! {\rm d}\ep\, \frac{\pa f(\ep-eV)}{\pa (eV)} (1+|r^A(\ep)|^2-|r^N(\ep)|^2),
\e
where $\ep$ is measured with respect to the Fermi energy and 
$f(\ep)$ is the Fermi distribution function 
for the temperature $T$. 
For $\ep\gg \Delta$ the Andreev amplitude $r^A$ vanishes and Eq.~(\ref{BTK}) reduces 
to the Landauer formula for the conductance.
For $\ep<\Delta$ the differential 
conductance in Eq.~(\ref{BTK}) is solely determined by $r^A$, since
$|r^N(\ep)|^2+|r^A(\ep)|^2=1$.

\begin{figure}[tb]
\centerline{\includegraphics[width=0.8\linewidth]{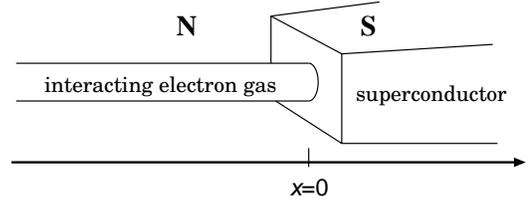}}
\caption{\label{fig:setup} 
The setup consisting of a one-dimensional weakly interacting electron gas for $x<0$
and a superconductor for $x>0$.  
}
\end{figure}

The Landauer formula 
as well as Eq.~(\ref{BTK}) play a key role in the scattering approach, which
captures the effects of geometry, boundaries, and disorder in nanoscopic samples, but 
ignores inelastic quasiparticle scattering.
In this Letter we apply Eq.~(\ref{BTK}) to the system depicted in Fig.~\ref{fig:setup}.
It consists of a superconductor in contact with a one-dimensional electron gas 
with a repulsive interaction, 
\be
\label{HI}
{\cal H}_{I}=\frac{1}{2}\s_{\sigma\sigma'}\!\int\!\!\!\int\! {\rm d}x\,{\rm d}y\,
\Psi\h_\sigma(x)\Psi\h_{\sigma'}(y)U_{x-y}\Psi_{\sigma'}(y)\Psi_{\sigma}(x),
\e
where $U_x$ is a symmetric positive-definite function of $x$, which  
is assumed to exponentially decay for $x$ exceeding some characteristic interaction range $d$. 

It is well-known that electron-electron interactions in a one-dimensional Fermi gas strongly modify 
the spectrum of low-energy excitations \cite{book}.  
In this case the elastic amplitudes $r^{N,A}(\ep)$ in (\ref{BTK}) have to be replaced by 
the renormalized ones $r_D^{N,A}(\ep)$ with the cutoff $D=\max\lt\{\ep,k_B T\rt\}$.
A way to calculate the renormalized scattering amplitudes
was developed in Refs.~\cite{MYG93,YGM94}, where the renormalization group (RG) equation
for the $S$-matrix of a single barrier in a one-dimensional interacting electron gas
was derived. The RG results are valid for arbitrary barrier transparencies 
but are restricted to the case of weak interactions. 
Complementary studies based on the bosonization technique \cite{KN92,FN93,VBBF02,FHO95,WFHS96}
are applicable for arbitrary interaction strengths in the limits
of very low or very high barrier transparency. The RG method of Refs.~\cite{MYG93,YGM94}
was subsequently extended to account for a resonant energy dependence of the bare $S$-matrix 
\cite{NG03,PG03}. We demonstrate below that the intrinsic 
energy dependence of Andreev reflection, which was disregarded in earlier works
\cite{TK97,FHOR99}, is crucial for understanding 
the transport properties of interacting normal-metal/superconductor nanostructures. 

A possible experimental realization of a quasi-one-dimensional electron system is provided by 
a single-wall carbon nanotube.
In recent experiments \cite{MKMD01} 
the differential Andreev conductance was dominated by the interface
between a carbon nanotube and a superconductor. 
A strong deviation from the predictions of the elastic theory was observed at low temperatures 
and low voltage bias, in which case the Andreev conductance was found to be strongly suppressed. 
The effect was shown to disappear with a slight change in the gate voltage applied to the
nanotube. The major role of the gate voltage in the experimental setup is to govern
the nanotube-superconductor coupling. Thus, 
the experiment suggests a strong sensitivity of the interaction-induced dip 
in the Andreev conductance to the normal-state resistance of the NS interface. 
 
Following the RG method of Refs.~\cite{NG03,PG03} we deal with an effective scattering matrix $S_D(\ep)$,
which acquires an additional dependence on the high-energy cutoff $D$. 
The renormalization procedure starts 
at a large $D=D_0\simeq \hbar v_{\rm F}/d$ with $v_{\rm F}$ the Fermi velocity.
The matrix $S_{D_0}$ coincides with the bare scattering matrix in the absence of interactions 
\be
\label{ini}
S_{D_0}(\ep)=S(\ep)=\lt(\ba{cc} r(\ep) & t'(\ep)\\ t(\ep) & r'(\ep) \ea \rt).
\e
The idea of the $S$-matrix renormalization is analogous to the poor man's scaling proposed 
by Anderson \cite{A70}. One starts with the analysis of
the first order in $U$ correction to the $S$-matrix, which is regularized by 
a formal truncation of the Fourier series of Hartree
and exchange potentials at the large momentum cutoff $D/\hbar v_{\rm F}$.  
The first order correction is, then, proportional to $\ln D/\ep$, showing
the logarithmic divergence in the limit $\ep\to 0$. 
By differentiating the first order result with respect to 
$D$ one arrives at the RG equation 
\be
\label{RG}
\frac{\pa S_D(\ep)}{\pa \ln D}=\Sigma_D- S_D(\ep) \Sigma_D\h S_D(\ep),
\e
which manifestly conserves the unitarity of the $S$-matrix \cite{LRS02}. 
The matrix $\Sigma_D$ depends on both $S_D(\ep)$ and the renormalized interaction constants \cite{Sol79}
\be
\label{g}
g_1=\frac{\alpha_1}{1-2\alpha_1\ln{(D/D_0)}},\qquad
g_2=\alpha_2+\frac{g_1-\alpha_1}{2},
\e
which makes Eq.~(\ref{RG}) a complex non-linear equation.
The bare values of the interaction constants
\be
\alpha_1=\int\!{\rm d}x\, \frac{U_x\,e^{2ik_{\rm F} x}}{2\pi\hbar v_{\rm F}}, \qquad
\alpha_2=\int\!{\rm d}x\, \frac{U_x}{2\pi\hbar v_{\rm F}},
\e
quantify backward and forward scattering, correspondingly. In order to guarantee
the validity of Eq.~(\ref{RG}) interactions are assumed to be weak, i.e. $\alpha_{1,2}\ll 1$.
The RG procedure is terminated at $D=\max\lt\{\ep, k_B T\rt\}$, so that
the renormalized scattering matrix at zero temperature is given by $S_{\ep}(\ep)$.

For a single resonant tunnel barrier in an interacting quantum wire
the RG equation for the $S$-matrix can be cast in the form of Eq.~(\ref{RG})
with \cite{NG03,PG03} 
\be
\label{tunnel}
\Sigma_D=\frac{1}{2}\lt(\ba{cc}(2g_1\!-\!g_2)\, r_{D}(-D)& 0 \\ 0 & (2 g'_1\!-\!g'_2)\, r'_{D}(-D)\ea\rt),
\e
where $g_{1,2}(D)$ and $g'_{1,2}(D)$ refer to the interaction constants 
on the left and the right side of the barrier, correspondingly. 
The matrix $\Sigma_D$ takes into account the contribution to the Friedel oscillation 
from electrons with energy $-D$ deep in the Fermi see. 
By lowering the cutoff according to Eq.~(\ref{RG}) the coherent backward scattering 
from the electron density oscillation in the entire energy range is included in the RG procedure.

In the presence of superconductivity the dimension of the $S$-matrix has to be doubled.
Therefore, the entries of $S(\ep)$ in the parametrization (\ref{ini}) 
have to be regarded as $2\times 2$ matrices in Nambu space, for example,
\be
\label{r}
r(\ep)=\lt(\ba{cc}r^N(\ep)& \bar{r}^A(\ep)\\
r^A(\ep)& \bar{r}^N(\ep) \ea\rt).
\e
The components of $r(\ep)$ fulfill an additional electron-hole symmetry
constraint $\sigma_y r(-\ep)^* \sigma_y =r(\ep)$,
where $\sigma_y$ is a Pauli matrix in Nambu space.

We find that the RG equation for the $S$-matrix of the NS interface
takes the form (\ref{RG}) with 
\be
\label{Sigma_NS}
\Sigma_D=\lt(\ba{cc}\sigma_D & 0\\ 0 & 0 \ea\rt),
\e
where $\sigma_D$ is the Nambu matrix
\beq
\label{sigmaD}
\sigma_D &=& \frac{1}{2}\lt(
\ba{cc}(2 g_1-g_2)r_D^N(-D) & (g_1+g_2) \bar{s}\\
(g_1+g_2) s & (2 g_1 - g_2)\bar{r}_D^N(D) 
\ea
\rt),\qquad\\
\label{s}
s&=&\frac{1}{2}\lt(r^A_D(-D)+r^A_D(D)\rt).
\eq
The lower diagonal block of $\Sigma_D$ vanishes due to the absence of renormalization
on the superconducting side of the interface. Another important difference from Eq.~(\ref{tunnel})
is the presence of the off-diagonal term (\ref{s}) in $\sigma_D$, 
which takes into account Friedel oscillations induced by Andreev reflection 
processes. Even though Andreev reflection can take place for energies above the 
gap, the combination $s$ is non-zero only for $D<\Delta$. Thus, 
the Andreev renormalization is effective only for subgap energies.

Equation (\ref{RG}) together with Eqs.~(\ref{Sigma_NS}-\ref{s}) is
the main result of the present work. 
The matrix $\sigma_D$ is obtained by differentiating 
the first order perturbation correction to the $S$-matrix
\be
\label{basis}
\sigma_D =\frac{\pa}{\pa\ln D}
\int\!\!\!\il_{\!\!\!\!-\infty}^{\!\!\!0}\!\frac{{\rm d}x\,{\rm d}y}{i\hbar v_{\rm F}}\, L_{k,x}
\tilde{U}_{x,y}(D) L_{k,y}, 
\e
in the limit $\ep_k\ll D$, where $\ep_k=\hbar v_{\rm F} k$. 
The transfer matrix $L_{k,x}$ 
describes free propagation of electron and hole 
quasiparticles 
\be
L_{k,x}=
\lt(\ba{cc}
e^{i(k_{\rm F}+k)x} & 0 \\
0 & e^{-i(k_{\rm F}-k)x} 
\ea\rt).
\e
The first order molecular potential $\tilde{U}_{x,y}(D)$ in Eq.~(\ref{basis})
depends on the cutoff and can be 
decomposed into 
Hartree and exchange terms $\tilde{U}_{x,y}=U^{H}_{x,y}-U^{ex}_{x,y}$,
that are given by, correspondingly,
\beq
\label{U}
U^{H}_{x,y}&=&\sigma_z\delta(x-y)\s_\sigma \int_{-\infty}^0\!\!\!{\rm d}z\, U_{x-z}
\la\Psi\h_\sigma(z)\Psi_\sigma(z)\ra,\\ \n
U^{ex}_{x,y}&=&\frac{1}{2}U_{x-y}\lt(\ba{cc}
2\la\Psi\h_\up(y)\Psi_\up(x)\ra & \la [\Psi_\up(x), \Psi_\down(y)]\ra\\
\la [\Psi\h_\down(x), \Psi\h_\up(y)]\ra &
-2\la\Psi\h_\down(x)\Psi_\down(y)\ra
\ea\rt),
\eq
where the square brackets denote the commutator.
The operators $\Psi$ in Eq.~(\ref{U}) have to be taken in the scattering basis
\be
\Psi_{\up}(x)=\int \frac{{\rm d}k}{\sqrt{2\pi}}\,
\lt( e^{i(k_{\rm F}+k)x}a_{\up k}+e^{-i(k_{\rm F}+k)x}b_{\up k}\rt),
\e
where the momentum integration is restricted 
to the interval $(-D/\hbar v_{\rm F}, D/\hbar v_{\rm F})$, leading to the 
cutoff dependence of $\tilde{U}_{x,y}$ in Eq.~(\ref{basis}). 
The operators $a_{\sigma k}$ and $b_{\sigma k}$ are related by the 
reflection matrix (\ref{r})
\be
\lt(\ba{c}b_{\up k}\\ b\h_{\down -k} \ea\rt)=
r(\ep_k) \lt(\ba{c}a_{\up k}\\ a\h_{\down -k} \ea\rt).
\e
The angular brackets in Eq.~(\ref{U}) are defined by
$\la a\h_{\sigma k} a^{\phantom{\dagger}}_{\sigma' k'}\ra =f(\ep_k) \delta_{\sigma\sigma'}\delta_{k k'}$.
Calculating $\sigma_D$ from Eq.~(\ref{basis}) in the limit $\ep_k, T \ll D$ 
we obtain the result (\ref{sigmaD}). 

Both the Hartree and exchange potentials in Eq.~(\ref{U})
are hermitian, $\tilde{U}_{y,x}\h=\tilde{U}_{x,y}$, and fulfill
an electron-hole symmetry, $\tilde{U}_{y,x}\h=\tilde{U}_{x,y}$,
for any $D$.
Thus, both the unitarity as well as the electron-hole symmetry 
of the $S$-matrix are conserved under the RG flow (\ref{RG}).

\begin{figure}[tb]
\centerline{\includegraphics[width=0.9\linewidth]{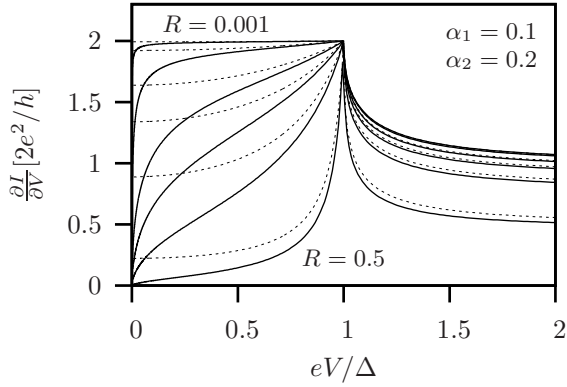}}
\caption{\label{figA} 
Zero temperature differential Andreev conductance versus the voltage bias
calculated from Eq.~(\ref{BTK}) with renormalized
reflection amplitudes. The solid curves result from the numerical solution of Eq.~(\ref{RG2})
with $D_0=100\Delta$ and correspond to different values of the normal-state resistance, 
$R=0.001, 0.01, 0.05, 0.1, 0.2, 0.5$, and to the particular choice 
of constants $\alpha_1=0.1$, $\alpha_2=0.2$. The dashed curves indicate 
the conductance in the absence of interactions. 
}
\end{figure}

The bare $S$-matrix for Andreev reflection has a specific energy dependence 
on the scale $\Delta$. For an ideal NS interface 
the bare Andreev reflection amplitude is given by  
\be
\label{gamma}
\gamma_\ep=\ep/\Delta-\sqrt{(\ep/\Delta)^2-1},\qquad \ep>0,
\e
where the positive branch of the square root has to be chosen for $\ep<\Delta$
and the relation $\gamma_{-\ep}=-\gamma^*_{\ep}$ extends the definition to negative energies.
Furthermore, the reflection matrix (\ref{r})
of a non-ideal interface can be conveniently parameterized as \cite{Bee97}
\be
\label{param}
r^N=e^{i\theta_1}\frac{(1-\gamma^2_\ep)\sqrt{R}}{1-\gamma^2_\ep R},\qquad
r^A=e^{i\theta_2}\frac{\gamma_\ep(1-R)}{1-\gamma^2_\ep R},
\e
where we assume the phases $\theta_{1,2}$ and
the normal-state reflection coefficient $R\in (0,1)$
to be energy independent.

Due to the simple structure of $\Sigma_D$ in Eq.~(\ref{Sigma_NS})
it suffices to consider the reflection block of Eq.~(\ref{RG}) 
\be
\label{RG2}
\frac{\pa r_D(\ep)}{\pa \ln D}=\sigma_D-r_D(\ep) \sigma_D\h r_D(\ep). 
\e
In what follows we focus on the physically relevant limit of short-range interactions 
$d < \xi$ or, equivalently, $D_0>\Delta$, where $\xi=\hbar v_{\rm F}/\Delta$ is the 
superconducting coherence length. We start from the analysis of Eq.~(\ref{RG2})
for the case of an ideal interface, $R=0$, and
parameterize $r^A_D(\ep)=\exp\lt(-i\phi_D(\ep)\rt)$, 
where the phase $\phi_D(\ep)$ is real for $\ep<\Delta$. 
The Andreev conductance below the gap is therefore not affected by interactions,
while the Andreev phase $\phi_D(\ep)$ 
strongly deviates from its elastic value $\arccos\ep/\Delta$.
Such an Andreev phase renormalization is  
absent in Refs.~\cite{TK97,FHOR99}.

With the substitution 
\be
\label{phi}
\phi_D(\ep)=\arccos
\frac{\ep/\Delta}{\cosh u_D -\sqrt{1-(\ep/\Delta)^2} \sinh u_D}
\e
Eq.~(\ref{RG2}) is reduced to the differential equation for a 
function of one variable 
\be
\label{u}
\frac{\pa u_D}{\pa \ln D}=(g_1(D)+g_2(D))\tanh(u_D-w_D),
\e
with the initial condition $u_\Delta=0$, and with the notation
$w_D= \arcth\sqrt{1-(D/\Delta)^2}$. The solution to Eq.~(\ref{u}) 
in the limit $D\ll \Delta$ can be approximated by
\be
u_D=(\alpha_2-\alpha_1/2)\ln\frac{\Delta}{D}
+\frac{3}{4}\ln\lt(1+a\ln\frac{\Delta}{D}\rt)+{\rm const.},
\e
where the parameter $a=2\alpha_1/(1+2\alpha_1\ln D_0/\Delta)$ 
has a logarithmic dependence on the initial cutoff $D_0\simeq \hbar v_{\rm_F}/d$.
From Eq.~(\ref{phi}) we obtain 
\be
\label{res}
\cos\phi_D(\ep)\simeq \frac{\ep}{\Delta}\lt(\frac{\Delta}{D}\rt)^{\alpha_2-\alpha_1/2}
\lt(1+a \ln \frac{\Delta}{D}\rt)^{3/4},
\e 
where $D=\max\{\ep,k_B T\}<\Delta$. At zero temperature we let $D=\ep$ in Eq.~(\ref{res})
and interpret the result as an additional suppression 
of the electron-hole coherence at finite $\ep$. 
The effect of the Andreev phase renormalization can be seen most explicitly 
in the supercurrent or in measurements of the density of states. 
A detailed analysis of these quantities, though, is beyond
the scope of the present work. We shall stress, however, that
the renormalization (\ref{res}) originates in the energy dependence 
of the bare scattering amplitudes, which has been ignored in the 
previous studies of the Josephson effect and of the density of states \cite{FHO95,WFHS96}.
  
For a non-ideal interface, $R\neq 0$, Eq.~(\ref{RG2}) 
reduces to a set of four coupled RG
equations for the variables $\gamma$, $R$ and $\theta_{1,2}$
in the parameterization (\ref{param}). It is remarkable that  
for energies above the gap only one equation remains
\be
\label{RD}
\frac{\pa R_D}{\pa \ln D}=(2g_1-g_2)R_D(1-R_D)\frac{1-\gamma_D^2}{1-\gamma_D^2R_D},
\e 
with the initial condition $R_{D_0}=R$ and with the function $\gamma_D$ given by the bare Andreev 
amplitude (\ref{gamma}). Both the amplitude $\gamma_\ep$ and the phases $\theta_{1,2}$
are not renormalized and do not 
acquire any $D$ dependence as far as $D>\Delta$.
It, then, follows from Eq.~(\ref{param}) that $r^N(\Delta)=0$, 
hence the differential Andreev conductance at $eV=\Delta$
equals its elastic value $4 e^2/h$. 
On the other hand, renormalization (\ref{RD}) reduces to the result of Refs.~\cite{MYG93,YGM94}
in the limit $\Delta\to 0$ and yields the well-known conductance suppression at low energies.
Thus the interactions tend to sharpen the non-monotonic behavior of the Andreev conductance
near $e V=\Delta$ in accordance with Ref.~\cite{LLYC03}. 

For energies below the gap, Eq.~(\ref{RG2}) is equivalent to a joint
renormalization of the three variables $\phi$, $R$ and $\theta_{1}$ in 
the parameterization (\ref{param}), 
where $\phi$ is the Andreev phase, $\gamma=\exp{(-i\phi)}$. 
Thus, below the gap the renormalization cannot be reduced to 
that of the single parameter $R$ in contradiction to  
the phenomenological description of Ref.~\cite{MKM}.
The phase $\theta_2$ is not renormalized in the entire energy range, 
which is related to the fact that it can be removed by an appropriate gauge transformation.
The Andreev conductance below the gap depends only on the absolute value 
of the amplitude $r^A_D(\ep)$. One can prove from the explicit form of the RG equations
that the energy dependence of $|r^A_D(\ep)|$, unlike that of $r^A_D(\ep)$, can be disregarded 
in the limit $\ep\ll \Delta$. Therefore, we obtain from Eq.~(\ref{RG2})
\cite{TK97,FHOR99}
\be
\label{TK}
\frac{\pa |r_D^A|^2}{\pa \ln D}=2(2\alpha_2-\alpha_1)|r^A_D|^2(1-|r_D^A|^2), \quad D\ll \Delta,
\e
which depends only on the invariant combination $2\alpha_2-\alpha_1$
of constants, which is positive for any repulsive interaction.
The general solution of Eq.~(\ref{TK}) can be written as
\be
\label{sol}
|r^A_D|^2=\frac{1}{1+b\lt(\Delta/D\rt)^{2(2\alpha_2-\alpha_1)}},
\e
where the coefficient $b$ has to be determined from the solution of Eq.~(\ref{RG2})
for $\ep,D\sim \Delta$. For a nearly ideal interface, $R\ll 1$, we find $b\propto R$,
which allows to estimate the width of the zero bias anomaly in the differential Andreev conductance as
$\ep_c \sim \Delta R^{1/(4\alpha_2-2\alpha_1)}$. 

We illustrate our findings in Fig.~\ref{figA}. The differential Andreev conductance 
at zero temperature is calculated from Eq.~(\ref{BTK}) with the renormalized amplitudes $r^{N,A}_\ep(\ep)$,
which are found by the numerical solution of Eq.~(\ref{RG2}). 
The interaction induced dip in the Andreev conductance
has a width $eV_c\sim \ep_c$ and is strongly sensitive to the normal-state resistance
of the interface. Such sensitivity was indeed observed in the experiments by 
Morpurgo {\it et al.} \cite{MKMD01}, where the normal-state transparency of the NS junction 
was controlled by means of a gate voltage applied to the nanotube. A small change in the gate voltage 
gave rise to a slight improvement of the normal-state transparency of the junction, which
had a drastic effect on the Andreev conductance. 

In conclusion we have derived and analyzed the energy-dependent RG equations for the 
scattering matrix of an interacting normal-metal/superconductor interface.
Our approach takes into account the intrinsic energy dependence of scattering
at the NS interface and is readily generalized to SNS structures
and to the case of a quasi-one-dimensional interacting normal metal. 
The effects of interaction on the Andreev conductance, but not on the Andreev phase, 
are shown to vanish in ideal NS junctions. Our results qualitatively 
explain recent experiments \cite{MKMD01} with carbon nanotubes.

We acknowledge D.~Aristov for drawing our attention to Ref.~\cite{LRS02}.
We also thank A.~Furusaki, L.~Glazman, and I.~V.~Gornyi for useful comments. 
This research was supported in part by the
NSF under Grant No. PHY99-07949, the Swiss NSF, the DFG
through SFB 513 and the BW-Research Network "Functional Nanostructures".

\end{document}